\documentclass[]{aa}
\usepackage{times}
\usepackage{natbib}
\usepackage[dvips]{graphicx}
\usepackage{epsfig,longtable,lscape}
\usepackage{epsfig}
\usepackage[figuresright]{rotating}
\bibpunct{(}{)}{;}{a}{}{,}

\usepackage{float}
\usepackage{amssymb}
\usepackage{amsmath}
\usepackage{aas_macros}
\usepackage{headings}
\usepackage{latexsym}
\usepackage{graphics}
\usepackage{graphicx}

\def\XMM{{\em XMM--Newton}}
\def\SMC{{\em SMC}}
\def\smc{{\em Small Magellanic Cloud}}

\def\EPIC{{\em EPIC}}

\def\MOS{{\em MOS}}

\def\pn{{\em pn}}

\def\ltsima{$\; \buildrel < \over \sim \;$}
\def\simlt{\lower.5ex\hbox{\ltsima}}
\def\gtsima{$\; \buildrel > \over \sim \;$}
\def\simgt{\lower.5ex\hbox{\gtsima}}

\begin{document}

\title{Highly absorbed X-ray binaries in the \smc}

\author{G. Novara\inst{1}, N. La Palombara\inst{1}, S. Mereghetti\inst{1}, F. Haberl\inst{2}, 
M. Coe\inst{3}, M. Filipovic\inst{4}, A. Udalski\inst{7}, A. Paizis\inst{1},  W. Pietsch\inst{2}, R. Sturm\inst{2}, M. Gilfanov\inst{5}, A. Tiengo\inst{1},  J. Payne\inst{4}, D. Smits\inst{6}, A. De Horta\inst{4}}

\institute{INAF-IASF, Istituto di Astrofisica Spaziale e Fisica Cosmica ``G.Occhialini'', Via Bassini 15, I--20133, Milano, Italy
 \and Max-Planck-Institut f{\" u}r extraterrestrische Physik, Giessenbachstra{\ss}e, 85748 Garching, Germany
 \and School of Physics and Astronomy, University of Southampton, SO17 1BJ, UK
 \and University of Western Sydney, Locked Bag 1797, Penrith South DC, NSW 1797, Australia
 \and Max-Planck-Institut f{\" u}r Astrophysik, Karl-Schwarzschild-Stra{\ss}e 1, 85741 Garching, Germany
 \and Department of Mathematical Sciences, University of South Africa, UNISA, Pretoria 0003, South Africa
 \and Warsaw University Observatory, Aleje Ujazdowskie 4, 00-478 Warsaw, Poland
}

\titlerunning{\XMM\, \smc}

\authorrunning{Novara et al.}

\offprints{Giovanni Novara, novara@iasf-milano.inaf.it}

\date{...}

\authorrunning{G. Novara et al.}

\titlerunning{Highly-absorbed X-ray binaries in the SMC}


\abstract{{Many of the high mass X-ray binaries (HMXRBs) discovered in recent years in
our Galaxy are characterized by a high absorption, most likely intrinsic to the system,
that can impede their detection at the softest X-ray energies. Exploiting the good
coverage  obtained with sensitive  \XMM\ observations,
we have undertaken a search for highly absorbed X-ray sources in the
Small Magellanic Cloud (SMC), which is known to contain a large number of HMXRBs.
After a systematic analysis of 62 \XMM\ SMC observations, we obtained a sample of 30
sources with evidence of an equivalent hydrogen column density larger
than 3$\times$$10^{23}$ cm$^{-2}$.}
{Five of these sources are clearly identified as HMXRBs, four being previously known (including
three X-ray pulsars) and one, XMMU J005605.8--720012,  being reported here for the first time.
For the latter, we present optical spectroscopy confirming the association with a Be star in the SMC.
The other sources in our sample have optical counterparts fainter than magnitude $\sim$16 in the V band,
and many have possible NIR counterparts consistent with highly reddened early-type
stars in the SMC.}
{While their number is broadly consistent with the expected population of background
highly absorbed active galactic nuclei, a few of them could be HMXRBs in which an early-type
companion is severely reddened by local material.}

}

\maketitle

\section{Introduction}

Many observations of the Small Magellanic Cloud (\SMC) in the X-ray energy
band have led to the discovery of a large number of High Mass X-ray Binaries (HMXRBs).
The number of  known HMXRBs in the \SMC\ (about one hundred, \citet{Liu2005}) is much larger than
expected by scaling the number of these sources seen
in the Milky Way according to the mass ratio of the two galaxies (M$_{MW}$/M$_{SMC}$ $\sim$50).
This has been interpreted as evidence of a recent episode of star
formation in the \SMC\ (\citet{Majid2004}, \citet{Shtykovskiy2007}, \citet{Antoniou2010}).
It is remarkable  that the \SMC\ contains only one supergiant system, SMC X-1.
All the other HMXRBs in the \SMC\ consist of neutron stars accreting from Be type companions.
Be X-ray binaries constitute also the largest class of HMXRBs in our Galaxy,
but the relative number of supergiant systems is much higher than in the \SMC,
since 30 of the 114 Galactic HMXRBs  \citep{Liu2006} are confirmed or suspected supergiant systems.

Many new Galactic HMXRBs have been discovered in the past few years with the
INTEGRAL satellite, thanks to 
high sensitivity in the hard X-ray range, coupled with an extensive monitoring of the Galactic plane. 
These observations led to the recognition of the new class of Supergiant Fast X-ray Transients
(SFXT) and to the discovery of persistent supergiant systems characterized
by a high absorption (equivalent column density above a few 10$^{23}$ cm$^{-2}$), which
escaped an earlier discovery because the high absorption severely suppresses their
flux  below 10 keV (see, e.g., \citet{Sidoli2010}, for a recent review).

To search for highly absorbed HMXRBs in the \SMC\, we carried out
a dedicated analysis of  the \XMM\ data  collected during  the large program for the \SMC\ 
survey performed by \citet{Haberl2008} and other observations of SMC targets.
In Sect.~\ref{sec:2}, we describe the X-ray observations and data
processing and in Sect.~\ref{sec:3} we describe the source detection and selection criteria
that led us to identify a sample of 30 highly absorbed sources.
Five of them are confirmed  HMXRBs, including four that were already known as X-ray binaries (Sect.~\ref{sec:4}).
In Sect.~\ref{sec:5} we present optical spectroscopy of the new source, demonstrating
that it is a Be system in the \SMC. We finally discuss our results in Sect.~\ref{sec:6}.

\section{X-ray observations and data processing}\label{sec:2}

We analyzed the 62 \XMM\ pointings of the \SMC\ performed from 2001 May 31 to 2010 March 16 listed in 
Table~\ref{tab:expo}.
The three \EPIC\ focal plane cameras \citep{Turner2001,Struder2001}
were operated in standard {\em Full Frame} mode.

The data of each camera were processed independently
using  the standard \XMM\ {\em Science Analysis Software} ({\em SAS}, v.10.0.0).
We first used the tasks {\tt emproc} and {\tt epproc} to obtain the \MOS\ and \pn\ event files of each
pointing.
We then filtered out time intervals affected by high instrumental background induced by
flares of soft protons (with energies below a few hundred keV).
To avoid contributions from genuine X-ray source variability,
this was done by examining  light curves binned at 100 s for   events with  energies above 10 keV and
PATTERN$\leq$4 or 0 for, respectively, the \MOS\ peripheral CCDs and the whole \pn\ camera;
the good time intervals (GTI) were selected by adopting  different count--rate
thresholds for the different pointings and instruments.
By selecting only events within GTI, we finally obtained three
``clean'' event lists for each observation (i.e. two for the \MOS\ and one for the \pn\ cameras),
whose exposure times are reported in Table~\ref{tab:expo}.
The total net exposure times are respectively 1572  and 1598 ks for the \MOS\ cameras,
and 1361 ks for the \pn\ one.

\begin{table*}
\begin{center}
\caption[]{Log of the \XMM\ observations of 62 \SMC\ fields with the corresponding net
        good time interval (GTI) for the two \MOS\ and the \pn\ cameras.}
\label{tab:expo}
\footnotesize{
\begin{tabular}{cccccc} \hline\noalign{\smallskip} \hline\noalign{\smallskip}
Observation ID & \XMM & Date & \multicolumn{3}{c}{GTI (ks)} \cr
       & revolution & (UT) & MOS1 & MOS2 & pn \cr
\noalign{\smallskip\hrule\smallskip}
 0011450101 & 0270 & 2001-05-31T02:20:10 &  42.8 & 43.0 & 40.8 \cr
 0011450201 & 0355 & 2001-11-16T03:23:32 &  37.2 & 38.5 & 40.1 \cr
 0018540101 & 0357 & 2001-11-20T23:42:37 &	26.7 & 27.0 & 23.4 \cr
 0084200101 & 0422 & 2002-03-30T13:48:28 &   8.8 & 8.2 & 7.1   \cr
 0084200801 & 0340 & 2001-10-17T10:07:40 &  20.7 & 20.7 & 17.1 \cr
 0110000101 & 0156 & 2000-10-15T15:18:28 &	25.2 & 25.7 & 14.7 \cr
 0110000301 & 0157 & 2000-10-17T21:45:54 &	12.6 & 12.4 & 7.4 \cr
 0112780201 & 0143 & 2000-09-19T02:05:37 &	3.6 & 2.3 & 0.0 \cr 
 0112780601 & 0254 & 2001-04-29T21:07:59 &	0.9 & 1.5 & 2.9 \cr
 0135721701 & 0721 & 2003-11-16T06:12:02 &	24.4 & 24.8 & 29.8 \cr
 0142660801 & 0721 & 2003-11-17T03:55:54 &	7.2 & 7.5 & 6.8 \cr
 0157960201 & 0737 & 2003-12-18T14:32:45 &	17.9 & 18.7 & 17.2 \cr
 0164560401 & 0803 & 2004-04-28T22:08:04 &	0.2 & 0.1 & 0.1 \cr
 0301170101 & 1151 & 2006-03-22T21:39:54 &  18.8 & 19.7 & 16.4 \cr
 0301170201 & 1151 & 2006-03-23T04:48:17 &  22.5 & 22.7 & 14.4 \cr
 0301170301 & 1158 & 2006-04-06T04:32:35 &  17.9 & 18.0 & 14.4 \cr
 0301170601 & 1153 & 2006-03-27T12:21:01 &  14.4 & 14.5 & 11.7 \cr
 0311590601 & 1146 & 2006-03-13T15:17:13 &	11.2 & 11.0 & 9.6 \cr
 0402000101 & 1248 & 2006-10-03T00:09:09 &	21.7 & 21.7 & 19.0 \cr
 0403970301 & 1329 & 2007-03-12T20:02:20 &  28.0 & 28.7 & 19.2 \cr
 0404680201 & 1263 & 2006-11-01T00:56:29 &  32.4 & 32.4 & 30.8 \cr
 0404680301 & 1344 & 2007-04-11T19:38:25 &  20.1 & 19.9 & 15.4 \cr
 0404680501 & 1344 & 2007-04-12T03:07:23 &  23.3 & 23.7 & 21.6 \cr
 0500980101 & 1380 & 2007-06-23T05:51:39 &  25.2 & 25.0 & 22.3 \cr
 0500980201 & 1372 & 2007-06-06T08:52:16 &  27.6 & 28.3 & 14.1 \cr
 0501470101 & 1371 & 2007-06-04T08:59:50 &	13.4 & 16.5 & 9.6 \cr
 0503000201 & 1444 & 2007-10-28T05:49:58 &  21.5 & 21.5 & 20.0 \cr
 0503000301 & 1514 & 2008-03-16T15:25:16 &  17.9 & 19.3 & 2.0  \cr
 0601210101 & 1727 & 2009-05-14T09:14:24 &  20.7 & 21.2 & 19.8 \cr
 0601210201 & 1794 & 2009-09-25T00:15:42 &  37.5 & 37.5 & 35.7 \cr
 0601210301 & 1729 & 2009-05-18T10:29:15 &  30.9 & 31.4 & 27.5 \cr
\hline\noalign{\smallskip} \hline\noalign{\smallskip}
\end{tabular}}
\end{center}
\end{table*}

\addtocounter{table}{-1}
\begin{table*}
\begin{center}
\caption[]{Continued}
\label{}
\footnotesize{
\begin{tabular}{cccccc} \hline\noalign{\smallskip} \hline\noalign{\smallskip}
Observation ID & \XMM & Date & \multicolumn{3}{c}{GTI (ks)} \cr
       & revolution & (UT) & MOS1 & MOS2 & pn \cr
\noalign{\smallskip\hrule\smallskip}
 0601210401 & 1794 & 2009-09-25T11:22:24 &  37.6 & 37.6 & 36.0 \cr
 0601210501 & 1794 & 2009-09-25T22:30:43 &  49.8 & 50.4 & 39.9 \cr
 0601210601 & 1795 & 2009-09-27T00:10:53 &  36.7 & 37.1 & 32.7 \cr
 0601210701 & 1795 & 2009-09-27T11:35:54 &  38.7 & 38.6 & 37.1 \cr
 0601210801 & 1801 & 2009-10-09T18:33:30 &  24.7 & 24.7 & 23.1 \cr
 0601210901 & 1795 & 2009-09-27T23:02:34 &  34.9 & 35.3 & 33.9 \cr
 0601211001 & 1817 & 2009-11-09T21:16:08 &  34.0 & 35.0 & 23.7 \cr
 0601211101 & 1806 & 2009-10-18T22:47:09 &  31.6 & 31.6 & 30.1 \cr
 0601211201 & 1807 & 2009-10-20T22:51:09 &  33.1 & 33.2 & 29.0 \cr
 0601211301 & 1798 & 2009-10-03T05:08:47 &  32.4 & 32.5 & 30.9 \cr
 0601211401 & 1814 & 2009-11-04T21:38:31 &  35.3 & 35.6 & 31.1 \cr
 0601211501 & 1803 & 2009-10-13T00:02:01 &  37.6 & 37.4 & 34.0 \cr
 0601211601 & 1802 & 2009-10-11T22:43:47 &  31.6 & 32.3 & 28.8 \cr
 0601211701 & 1804 & 2009-10-16T01:05:21 &  25.1 & 26.6 & 18.4 \cr
 0601211801 & 1819 & 2009-11-13T20:59:56 &  26.2 & 30.2 & 22.3 \cr
 0601211901 & 1827 & 2009-11-30T14:46:19 &  31.7 & 31.7 & 30.1 \cr
 0601212001 & 1826 & 2009-11-27T22:39:06 &  28.3 & 28.3 & 26.3 \cr
 0601212101 & 1820 & 2009-11-16T05:48:08 &  34.1 & 34.1 & 32.5 \cr
 0601212201 & 1822 & 2009-11-16T06:10:27 &  27.1 & 29.4 & 24.3 \cr
 0601212301 & 1786 & 2009-09-09T09:13:53 &  33.5 & 33.5 & 31.9 \cr
 0601212401 & 1750 & 2009-06-29T14:46:19 &  29.5 & 30.9 & 23.9 \cr
 0601212501 & 1786 & 2009-09-09T19:13:54 &  33.5 & 33.5 & 31.9 \cr
 0601212601 & 1750 & 2009-06-29T06:04:39 &  26.0 & 27.1 & 17.5 \cr
 0601212701 & 1840 & 2009-12-26T07:25:22 &  36.7 & 36.7 & 34.5 \cr
 0601212801 & 1831 & 2009-12-07T23:35:54 &  21.3 & 21.3 & 16.7 \cr
 0601212901 & 1788 & 2009-09-13T13:29:26 &  36.1 & 36.1 & 34.5 \cr
 0601213001 & 1788 & 2009-09-13T01:11:03 &  41.6 & 41.7 & 38.9 \cr
 0601213201 & 1878 & 2010-03-12T00:56:15 &  12.5 & 13.2 & 9.4  \cr
 0601213301 & 1878 & 2010-03-12T05:26:15 &   9.8 & 9.9  & 8.1  \cr
 0601213401 & 1880 & 2010-03-16T10:05:12 &  17.6 & 17.5 & 11.9 \cr
 0656780201 & 1886 & 2010-03-27T12:20:41 &	11.0 & 12.3 & 7.0 \cr
\hline\noalign{\smallskip} \hline\noalign{\smallskip}
\end{tabular}}
\end{center}
\end{table*}

\section{Source detection and selection criteria}\label{sec:3}

\begin{figure}[h!]
\centering
\includegraphics[width=6.5 cm, angle=-90]{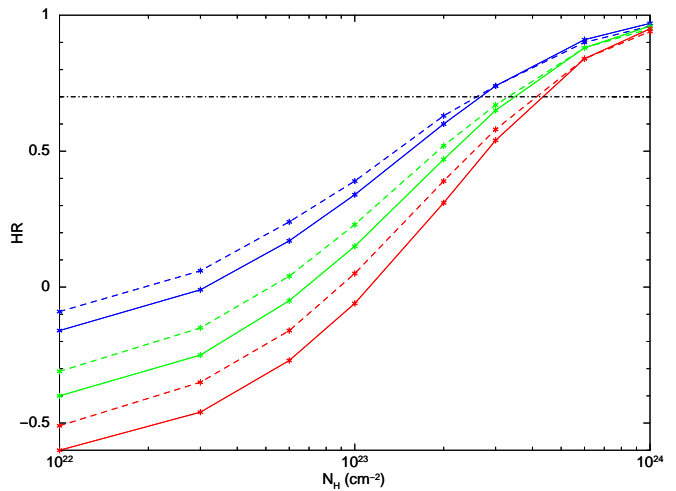}
\caption{Expected hardness ratio (of the count
rates in the energy ranges 1--3 to 3--10 keV) as a
function of the column density. Solid and dashed lines
refer to the \pn\ and the \MOS\ cameras, respectively.
Red, green, and blue lines refer to power-law spectra
with photon indices of 2, 1.5, and 1, respectively.
The horizontal line at HR = 0.7 represents the threshold
value used to select highly absorbed sources.}
\label{NH_HR}
\end{figure}

We optimized our source detection strategy to select highly absorbed X-ray sources.
To this aim, we simulated the expected \EPIC\ count distributions for absorbed power--law spectra,
with photon--index $\Gamma$ in the range 1--2.
We used two absorption components: the first one, with column density fixed at
$6\times10^{20}$ cm$^{-2}$ and elemental abundances from \citet{Wilms2000}, accounts for the
foreground absorption in our Galaxy;
for the second one, which accounts for the absorption in the SMC and is intrinsic to the source, we
considered  N$_{\rm H}$ values in the range 10$^{22}$--10$^{24}$ cm$^{-2}$,
and metal abundances of 0.2, as is typical of the SMC \citep{Russell1992}.
We defined the \textit{Hardness Ratio} HR = $(H-S)/(H+S)$),
based on the count rates in the soft (S, 1--3 keV) and hard (H, 3--10 keV) energy ranges.
These simulations showed that, for $\Gamma$ = 2 and N$_{\rm H} > 3\times10^{23}$ cm$^{-2}$,
the fraction of the detected source counts with energies above 3 keV is at least 70 \%,
and that this fraction is even higher for lower values of $\Gamma$.
In Fig.~\ref{NH_HR}, we report the estimated HR values as a function of N$_{\rm H}$,
for both the \pn\ and \MOS\ cameras and different values of $\Gamma$.
On the basis of these simulations, we adopted a HR threshold value of 0.7,
in order  to select sources with N$_{\rm H} > 3\times10^{23}$ cm$^{-2}$.

For each observation in Table~\ref{tab:expo}, we used the cleaned event
file to produce merged images from the three cameras in the 1--3 and 3--10 keV energy ranges,
according to the procedure described in detail in \citet{Baldi2002}.
For the 3--10 keV  pn images, we excluded the energy range 7.8--8.2 keV,
in order to reduce the instrumental background caused by the Cu line.
For each image, we also produced the corresponding merged exposure map, which
accounts for 
mirror vignetting and effective field of view.
The total count-rate-to-flux conversion factors were obtained as a mean of the
\textit{MOS1/MOS2/pn} factors, weighted on the individual exposures of the three
cameras. The source detection was based on a maximum likelihood technique, as described
in detail in \citet{Novara2006}, and used background maps produced
with the correction  algorithm described in \citet{Baldi2002}.
For each energy band, we only selected sources with a final detection likelihood {\em -ln~P} $>$ 8.5
(where {\em P} is the probability of a false
detection due to a Poissonian random fluctuation of the background).
This likelihood threshold corresponds to a $\sim$3 $\sigma$ detection.
After manually removing a few spurious detections caused, for instance, by events falling close
to the CCD edges, we obtained a final master list  containing $\simeq$1500 sources.
For each source, the master list provides various
parameters including the detector and sky coordinates, the effective
exposure time, the total counts, count--rate and errors in the different energy ranges,
and the detection likelihood.

\begin{landscape}

\begin{table}

\caption{Main characteristics of the highly absorbed X-ray sources.}\label{sources}
\begin{tabular}{ccccccrccl} \hline \hline
(1)	& (2)			& (3)		& (4)	& (5)	& (6)		& (7)		& (8)		& (9)		& (10)	\\
SRC	& NAME			&  RA		& DEC	& ERROR	& Likelihood	&  RATE		& f$_{X}$	& HR 		& Remarks	 \\
	&		& h:m:s	& $\circ$:$\prime$:$\prime\prime$ & arcsec & &  ($\times 10^{-4}$ cts s$^{-1}$)	& erg cm$^{-2}$ s$^{-1}$	 &        &	\\ \hline
01	& XMMUJ004102.6--732530 & 00:41:02.69 & --73:25:30.9 & 2.1  &  27.43 & 13.6 $\pm$ 2.9  & 8.94E--14  & 0.86  $\pm$  0.23   &	\\
02	& XMMUJ004210.9--733011 & 00:42:10.90 & --73:30:11.8 & 2.0  &  16.80 & 12.4 $\pm$ 3.3  & 8.21E--14  & 1.00  $\pm$  0.29   &	\\
03	& XMMUJ004226.3--730417 & 00:42:26.38 & --73:04:17.4 & 1.5  &  37.07 & 11.7 $\pm$ 2.0  & 7.76E--14  & 0.73  $\pm$  0.20   &	\\
04	& XMMUJ004244.7--732356 & 00:42:44.77 & --73:23:56.4 & 2.1  &  15.34 & 5.9  $\pm$ 1.6  & 3.89E--14  & 0.80  $\pm$  0.34   &	\\
05	& XMMUJ004536.2--724131 & 00:45:36.23 & --72:41:31.3 & 1.7  &  10.74 & 18.4 $\pm$ 5.8  & 1.18E--13  & 0.73  $\pm$  0.29   &	\\
06	& XMMUJ004814.0--731006 & 00:48:14.07 & --73:10:06.0 & 1.5  &  58.28 & 55.4 $\pm$ 8.2  & 3.49E--13  & 0.73  $\pm$  0.12   &	XMMUJ004814.1-731003, SXP25.5 (1)	\\
07	& XMMUJ004818.7--732102 & 00:48:18.74 & --73:21:02.4 & 1.9  &  $>$86 & 103  $\pm$ 9.1  & 6.81E--13  & 0.81  $\pm$  0.08   &	[SG2005] SMC 34 (2)			\\
08	& XMMUJ004853.5--732455 & 00:48:53.51 & --73:24:55.7 & 2.0  &  $>$86 & 84.3 $\pm$ 8.2  & 5.71E--13  & 0.72  $\pm$  0.07   &						\\
09	& XMMUJ004911.5--731717 & 00:49:11.57 & --73:17:17.8 & 1.8  &  12.79 & 15.1 $\pm$ 4.8  & 9.86E--14  & 0.91  $\pm$  0.34   &	CXOU J004910.7-731717 (3)		\\
10	& XMMUJ005020.7--720907 & 00:50:20.74 & --72:09:07.0 & 0.6  &  $>$86 & 68.6 $\pm$ 5.3  & 4.51E--13  & 0.94  $\pm$  0.05   &						\\
11	& XMMUJ005306.6--722400 & 00:53:06.66 & --72:24:00.2 & 1.5  &  12.96 & 6.0  $\pm$ 1.8  & 3.93E--14  & 0.72  $\pm$  0.33   &						\\
12	& XMMUJ005322.4--715927 & 00:53:22.44 & --71:59:27.5 & 1.0  &  10.35 & 11.2 $\pm$ 3.6  & 7.41E--14  & 0.87  $\pm$  0.36   &						\\
13	& XMMUJ005605.8--720012 & 00:56:05.85 & --72:00:12.2 & 1.6  &  26.57 & 8.4  $\pm$ 1.7  & 5.51E--14  & 0.70  $\pm$  0.23   &	New Be binary				\\
14	& XMMUJ005722.4--713114 & 00:57:22.44 & --71:31:14.8 & 0.9  &  12.66 & 15.7 $\pm$ 4.3  & 1.03E--13  & 0.72  $\pm$  0.33   &						\\
15	& XMMUJ005724.3--715917 & 00:57:24.31 & --71:59:17.8 & 1.5  &  16.81 & 17.8 $\pm$ 4.6  & 1.20E--13  & 0.84  $\pm$  0.29   &						\\
16	& XMMUJ005732.5--712926 & 00:57:32.51 & --71:29:26.3 & 1.8  &  8.93  & 11.3 $\pm$ 3.7  & 7.48E--14  & 0.86  $\pm$  0.46   &						\\
17	& XMMUJ005735.8--721935 & 00:57:35.82 & --72:19:35.1 & 2.0  &  30.99 & 21.4 $\pm$ 4.5  & 1.45E--13  & 1.00  $\pm$  0.18   &	XMMU J005735.7-721932 (4)		\\
	&			&		&	&	&	&			&	&			&	CXOU J005736.2-721934, SXP565 (5)	\\
18	& XMMUJ005812.9--723049 & 00:58:12.92 & --72:30:49.5 & 1.4  &  $>$86 & 1330 $\pm$ 30   & 8.78E--12  & 0.77  $\pm$  0.01   &	RX J0058.2-7231, SXP293 (6)		\\
19	& XMMUJ010115.2--721640 & 01:01:15.28 & --72:16:40.6 & 2.1  &  13.14 & 19.7 $\pm$ 5.5  & 1.30E--13  & 0.82  $\pm$  0.35   &						\\      
20	& XMMUJ010233.9--723443 & 01:02:33.99 & --72:34:43.2 & 1.8  &  60.86 & 29.7 $\pm$ 4.9  & 1.90E--13  & 0.91  $\pm$  0.17   &						\\
21	& XMMUJ010248.6--730822 & 01:02:48.69 & --73:08:22.6 & 1.4  &  39.22 & 28.2 $\pm$ 6.2  & 1.83E--13  & 0.77  $\pm$  0.23   &						\\
21	& XMMUJ010432.5--722543 & 01:04:32.56 & --72:25:43.6 & 2.1  &  12.73 & 8.9  $\pm$ 2.5  & 5.81E--14  & 0.85  $\pm$  0.36   &						\\
23	& XMMUJ010802.9--722627 & 01:08:02.93 & --72:26:27.5 & 2.1  &  17.50 & 7.9  $\pm$ 2.2  & 5.21E--14  & 0.71  $\pm$  0.30   &						\\
24	& XMMUJ010811.9--721005 & 01:08:11.91 & --72:10:05.2 & 1.2  &  24.42 & 7.0  $\pm$ 1.6  & 4.61E--14  & 0.81  $\pm$  0.27   &						\\
25	& XMMUJ010831.4--730630 & 01:08:31.40 & --73:06:30.9 & 1.6  &  39.61 & 12.1 $\pm$ 2.1  & 7.98E--14  & 0.93  $\pm$  0.19   &						\\
26	& XMMUJ010842.6--723839 & 01:08:42.68 & --72:38:39.6 & 1.8  &  25.94 & 39.5 $\pm$ 8.0  & 2.56E--13  & 0.79  $\pm$  0.17   &						\\
27	& XMMUJ010908.0--720642 & 01:09:08.09 & --72:06:42.1 & 1.3  &  13.52 & 6.3  $\pm$ 1.9  & 4.11E--14  & 0.85  $\pm$  0.35   &						\\
28	& XMMUJ011339.7--724733 & 01:13:39.74 & --72:47:33.3 & 2.0  &  11.87 & 15.0 $\pm$ 4.8  & 1.01E--13  & 0.86  $\pm$  0.31   &						\\
29	& XMMUJ011434.8--730730 & 01:14:34.83 & --73:07:30.4 & 2.2  &  10.66 & 14.7 $\pm$ 4.7  & 9.70E--14  & 0.70  $\pm$  0.29   &						\\
30	& XMMUJ011807.8--731713 & 01:18:07.84 & --73:17:13.9 & 2.2  &  16.17 & 24.6 $\pm$ 6.3  & 1.60E--13  & 0.85  $\pm$  0.23   &						\\
\hline \hline
\end{tabular}
\vfill
Key to Table: Col.(1) = source ID; Col.(2) = source catalogue name; Col.(3) = right ascension (J2000); Col.(4) = declination (J2000); 
Col.(5) = position error (1 $\sigma$ c.l.); Col.(6) = detection likelihood; Col.(7) = 3--10 keV count rate 
(sum of the PN and MOS counts divided by the sum of PN and MOS exposures); Col.(8) = X-ray flux in the energy band 3--10 keV, assuming a power--law emission model with a hydrogen column density N$_{\rm H} > 3\times10^{23}$ cm$^{-2}$ and a photon--index $\Gamma$ = 1.5; Col.(9) = hardness ratio between energy bands 1--3 keV and 3--10 keV; Col.(10) = likely identification with already known sources. References are given in parenthesis:

(1) \citet{Haberl2008b}
(2) \citet{Shtykovskiy2005}
(3) \citet{Laycock2010}
(4) \citet{Sasaki2003}, \citet{Haberl2008a}
(5) \citet{Macomb2003}
(6) \citet{Haberl2008a}
(7) \citet{McGowan2008}

\end{table}
\end{landscape}

\begin{table*}
\begin{center}
\caption{Candidate optical-IR counterparts of the highly absorbed X-ray sources.}\label{counterparts}
\begin{tabular}{cclrrcc} \hline \hline
(1)	& (2)				& (3)		& (4)			& (5)	  & (6)	   & (7)     \\
ID	& NAME				& V		& f$_{X}$/f$_{V}$	& K	  & B--V  & J--K     \\
	&				& mag   	& log$_{10}$		& mag 	  & mag    & mag     \\ \hline
01	& IRSF 00410291-7325332 & 20.79$^{(a)}$ & 0.76	&  16.68  &  1.21  &  1.61   \\
02	& IRSF 00421082-7330127 & 19.31$^{(a)}$ & 0.13	&  17.00  &  1.05  &  0.63   \\
03	& IRSF 00422638-7304184 & $>$23 & $>$1.58	&  15.65  &  -     &  2.02   \\
04	& IRSF 00424550-7323553 & 17.05 	& -1.09	&  13.90  &  1.39  &  0.87   \\
05	& IRSF 00453646-7241343 & 19.43 	&  0.34	&  15.72  &  0.73  &  1.84   \\
06	& IRSF 00481410-7310040 & 15.30	& -0.85 &  15.64  &  0.26  &  0.04   \\
07	& IRSF 00481871-7321000 & 16.18 	& -0.21	&  15.29  &  0.25  &  0.36   \\
08	& IRSF 00485330-7324574 & $>$23$^{(a)}$ & $>$2.44 & 15.55 &  -    &  1.30   \\
09	& IRSF 00491089-7317172 & 18.80	&  0.01	&  15.95  &  1.00  &  0.38  \\
10	& IRSF 00502062-7209073 & $>$23	&  2.34	&  16.86  &  -     &  1.65  \\
11	& ? & $>$23  & $>$1.28			&  $>$17  &   -    &  -     \\
12	& MCPS 2493874		 & 20.74	&  0.65	&  $>$17  &  0.47  &  -     \\
13	& IRSF 00560575-7200118 & 16.71 & -1.08	&  17.18  &  -0.11 &  -0.11 \\
14	& ? & $>$23  & $>$1.70			&   $>$17   &   -    &  -     \\
15	& MCPS 3076421		 & 21.42 &  1.14	&  $>$17  &  0.35  &  -     \\
16	& ? & $>$23  & $>$1.56			&    $>$17    &   -    &  -     \\
17	& IRSF 00573600-7219342 & 15.99 & -0.95	&  15.35  &  0.01  &  0.42  \\
18	& IRSF 00581258-7230488 & 14.87 &  0.38	&  14.14  &  0.10  &  0.39  \\
19  & IRSF 01011529-7216374 & 19.23$^{(a)}$ &  0.30 & 14.69  &  1.28  &  1.59  \\ 
20	& IRSF 01023342-7234421 & 19.64$^{(a)}$ &  0.62 & 16.09  &  1.14  &  1.68  \\
21	& IRSF 01024839-7308213 & 20.99 	 &  1.15 & 17.08  &  1.04  &  1.90  \\
22	& IRSF 01043305-7225407 & 16.23 	& -1.25 &  13.12  &  1.38  &  0.81  \\
23	& MCPS 4411974		 & 20.01  	& 0.21	&  $>$17  &  0.04  &  -     \\
24	& IRSF 01081153-7210048 & $>$23  & $>$1.35	&  17.26  &   -    &  1.61  \\
25	& IRSF 01083147-7306325 & 18.00  &  -0.41	&  15.84  &  0.98  &  0.50  \\
26	& IRSF 01084310-7238375 & 19.53  &  0.71	&  17.84  &  0.84  &  0.21  \\
27	& IRSF 01090778-7206438 & 21.13  &  0.55	&  $>$17  &  0.57  &  -     \\
28	& MCPS 4879163		 & 22.36  &  1.44	&  $>$17  &  0.56  &  -     \\
29	& ? & $>$23  & $>$1.67			&  $>$17  &   -    &  -     \\
30	& ? & $>$23  & $>$1.89			&  $>$17  &   -    &  -     \\
\hline \hline
\end{tabular}
\vfill
Key to Table: Col.(1) = source ID; Col.(2) = name of the brightest infrared (IRFS sources, \citep{Kato2007}) or optical (MCPS sources, \citep{Zaritsky2002}) candidate counterpart within the X--ray error circle (a `?' is reported when no counterpart is found); Col.(3) = {\em V} magnitude; Col.(4) = X-ray--to--optical flux ratio (in log$_{10}$ units); Col.(5) = {\em K} magnitude; Col.(6) = {\em B--V} color index; Col.(7) = {\em J--K} color index

Notes
$^{(a)}$ for this object there is a possible counterpart with a slightly brighter V magnitude.

\end{center}
\end{table*}

\section{Results}\label{sec:4}

 \begin{figure}[h!]
 \centering
 \includegraphics[width=6.5 cm, angle=-90]{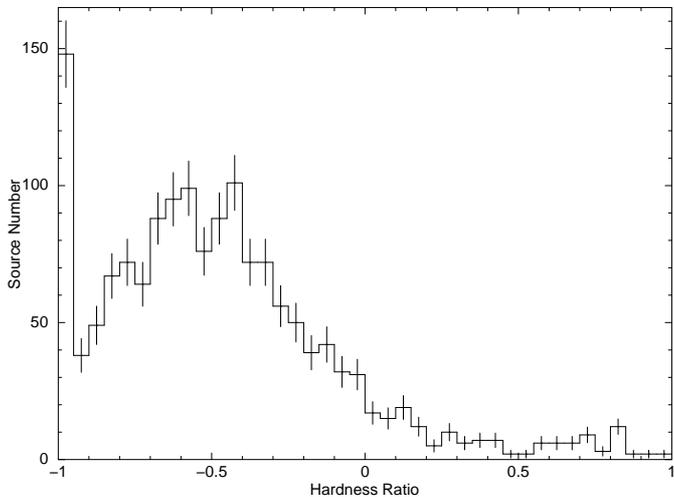}
 \caption{Hardness-ratio distribution of the detected X--ray sources}
 \label{HR}
 \end{figure}

In Fig.~\ref{HR}, we report the HR distribution of all the detected sources. This figure illustrates that most sources have
HR $<$ 0, with a peak around -0.5, hence they are characterized by non-absorbed soft spectra. On the
other hand, only very few sources have high HR values.
In our final source list, there are 30 objects with HR $>$ 0.7, which
we consider as highly absorbed sources with N$_{\rm H} > 3 \times 10^{23}$ cm$^{-2}$;
given the HR distribution, we expect that, even if we take into account the HR errors, only a very few
additional sources would have HR $>$ 0.7.
The sky coordinates of these 30 objects, corrected for the satellite pointing uncertainty,
are listed in Table \ref{sources}. The astrometric correction of the X-ray images was done
by cross--matching the brightest sources of each field with the Magellanic Clouds Photometric Survey optical
catalogue (MCPS, \citet{Zaritsky2002}), and selecting the X-ray sources with a single,
clearly evident optical counterpart within a 3$^{\prime\prime}$ error radius.
We then used the IRAF task {\tt geomap} to compute the linear transformation between the
X-ray and optical coordinates of these reference sources.
The calculated X-ray--to--optical coordinate transformations have rms fit residuals between 0\farcs43 and 1\farcs85 at 1 $\sigma$ c.l.,
with an average value of 1\farcs1; they correspond to the (radial) systematic astrometric errors in the corrected observations.
These transformations were then used by the IRAF task {\tt geoxytran} to obtain the corrected X-ray coordinates given in Table~\ref{sources}.
The total uncertainty on the X-ray  position of each source is given by the sum in quadrature of
the calculated systematic astrometric
error in the observation and the measured statistical error of the source itself.

We also list in Table~\ref{sources} the hardness ratios, count rates, and fluxes of the sources.
The fluxes refer to the 3--10 keV energy range, and have been computed assuming a
power-law with photon index $\Gamma$ = 1.5 and N$_{\rm H}$ = 3$\times$10$^{23}$ cm$^{-2}$.
The brightest source has a absorbed flux of  9$\times10^{-12}$ erg cm$^{-2}$ s$^{-1}$, 
while the faintest sources in our sample have fluxes of $\sim$4$\times10^{-14}$ erg cm$^{-2}$ s$^{-1}$ corresponding to  $\sim$2$\times10^{34}$ erg s$^{-1}$,
assuming a \SMC\ distance of 60 kpc \citep{Hilditch2005}

Some of our selected sources were already reported in other studies of the \SMC,
in these cases we give in the table their original name and the relevant references.
Four sources, including three pulsars, were already known as HMXRBs:  
SXP25.5, SMC34, SXP565, and SXP293.
We searched for possible counterparts of our highly absorbed sources using the MCPS catalogue,
which is based on  observations performed in the {\em U}, {\em B}, {\em V}, and {\em I} filters
between November 1996 and December 1999 \citep{Zaritsky2002},
and the InfraRed Survey Facility (IRSF) Magellanic Cloud catalogue \citep{Kato2007} for the  NIR   
({\em J}, {\em H }, and {\em K}  filters).

For each source, we report in Table \ref{counterparts} the optical and NIR properties of the
brightest object in the K band present within a distance equal to two times
the 1 $\sigma$ position error (with only five exceptions, marked in the table, 
the brightest object in K is also the brightest in V). 
For five sources, no catalogued optical or NIR counterparts were found
to limits as faint as of V$\sim$23 mag and K$\sim$17 mag. 

\begin{figure}[h!]
\centering
\includegraphics[width=6.5 cm, angle=-90]{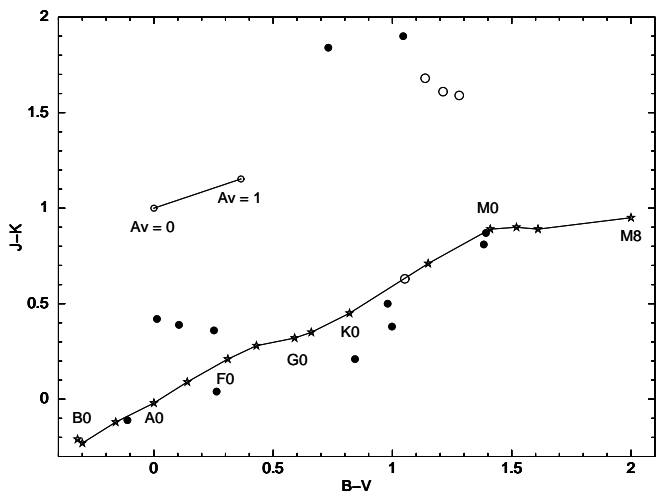}
\caption{Color indicies of the candidate counterparts of the 
highly absorbed X-ray sources of our sample. 
For the sources reported with a filled circle, the brightest optical and infrared candidate counterparts are coincident, 
while the sources reported as an open circle correspond to those with the `a' flag in Table~\ref{counterparts}: 
in this case, there is also a possible alternative counterpart, with a slightly brighter V magnitude, to the selected infrared counterpart. 
We also indicate the main-sequence stars and a bar which shows the reddening corresponding to A$_V$=1 mag.}
\label{colors}
\end{figure}

A plot of the  NIR versus optical colors shows that the four already known 
HMXRBs of our sample are located in the region corresponding to J--K$<$0.5 mag and B--V$<$0.5 mag (see Fig. \ref{colors}). 
As discussed in Sect.~5, the other source falling in this region (source \# 13) is 
a new Be binary system in the SMC. Three of these sources are well above the position of the main--sequence stars 
(whose colors are taken from \citet{Johnson1966}): if they are affected by an IR excess (which could be due to the decretion disk 
of the Be star), their optical/NIR colors are consistent with early-type stars with moderate optical reddenings (A$_V\sim$ 2--3 mag).
On the other hand, HMXRBs in which the optical companion is also affected by a high local absorption would show much redder 
colors, as seen in several of the new HMXRBs discovered by INTEGRAL in our Galaxy. Therefore, we cannot exclude that some of the 
other sources plotted in Fig. \ref{colors}, or sources for which no optical and NIR counterparts have been detected, are also highly absorbed HMXRBs. 
The five sources at the top of the plot are far from the main--sequence stars, therefore it is difficult to suggest a reliable 
classification of them; this is particularly true for the three sources shown as open circles, for which a different 
counterpart could also be suggested. These five sources could be early-type stars only if they are characterized by both a large 
IR excess and a high reddening. On the other hand, since for AGNs J--K $>$ 0.5 \citep{Kouzuma2010} and -0.4 $<$ B--V $<$ 0.7 \citep{Hatziminaoglou2002}, 
these sources could be more probably classified as reddened AGNs.

To search for possible periodicities in any optical counterparts close to the \XMM\ sources, data from
OGLE III were acquired for all objects within 4$^{\prime\prime}$ of the 30 \XMM\ position.
We applied a Lomb-Scargle analysis to a total of 129 lightcurves, searching for periods in the range 2--200 days. 
Periods shorter than
2 days were avoided because they are getting close to the average daily sampling of the OGLE data, and periods
longer than 200 days approach the annual sampling patterns. In each case, the data were first de-trended using a
polynomial of order three before being searched. In some cases, there were many optical counterparts;
conversely, other sources had no optical objects within the search zone.

Significant periodicities were found for four sources (three of which were already known) and tentative
periods for an additional three objects (see Table~\ref{ogle}). That three previously known periods were
``re-discovered'' in this search reassures us that the techniques used are effective in identifying
possible periods. The one new significant detection (source number 19) shows a modulation that is extremely
sinusoidal in nature when the data are folded at the period of 199.6 days. This modulation shape is not normal
for a long-period binary modulation, which tends to reveal short outbursts coincident with the periastron
passage of the neutron star (see, for example, \citet{Coe2004}). Such smooth sinusoidal modulations tend to
be associated with pulsations in stars. A likely possibility is that it could be an Non-Radial Pulsations (NRP) in a B-type star with
a period very close to the one day sampling, with the result that the signal at 199.6 days is the beat frequency of
the true pulse period and the sampling period.

\begin{table*}
\begin{center}
\caption{List of OGLE III objects within 4'' of an \XMM\ position that show evidence of a periodic signal}
\label{ogle}
\begin{tabular}{ccccccc}\hline \hline

XMM ID & OGLE III ID & V & I & LS power & Period & Comments \\ \hline

2 & 22605 & 21.3 & 20.3 & 12 & 2.16 days & Tentative \\
6 & 50768 & 15.8 & 15.7 & 14 & 22.5 days & SXP25.5 \citep{Rajoelimanana2011} \\
8 & 20669 & ?    & 21.0 & 13 & 4.08 days & Tentative \\
17& 27992 & 16.1 & 15.8 & 26 & 157.2 days & SXP565 \citep{Bird2011} \\
18& 19191 & 14.9 & 14.6 & 200& 59.7 days & SXP293 \citep{Schmidtke2004} \\
20& 43690 & 20.8 & 21.0 & 15 & 5.96 days & Tentative \\
26& 22676 & 19.6 & 18.6 & 23 & 199.6 days & Possible new period \\
\hline \hline
\end{tabular}
\end{center}
\end{table*}

\section{XMMU J005605.8--720012: A new Be HMXRB in the \SMC}\label{sec:5}

Source \# 13 was detected with a total of 68 counts in the
hard energy range 3--10 keV and a HR value of 0.71, just above our threshold.
It has a relatively bright ($V$ = 16.71 mag) optical counterpart, with color B--V = --0.12 mag,
and with a quality flag in the MCPS indicating a successful fit with a stellar atmosphere model.
It has an unabsorbed luminosity, in the energy range 3--10 keV, of 4.4$\times$10$^{34}$ erg s$^{-1}$.

\begin{figure}[h!]
\centering
\includegraphics[width=9.5 cm, angle=0]{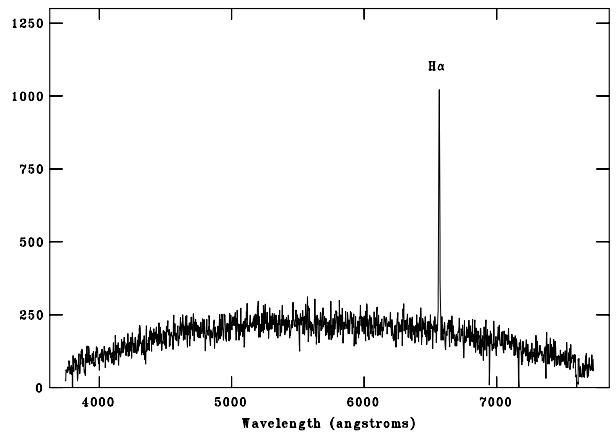}
\caption{SAAO optical spectrum of source \# 13; note the presence of a strong H$_{\alpha}$ emission line.}
\label{Halpha}
\end{figure}

Optical spectral observations of source \# 13 were obtained in September 2010, using the
1.9-meter telescope and the Cassegrain spectrograph at the South African Astronomical
Observatory (SAAO) in Sutherland. We used grating number 7 (300 lines per mm) to obtain
spectra between 3700 \AA\ and 7700 \AA\ at a resolution of 5 \AA. For these, the slit size
was 1\farcs5 $\times$ 1\farcm5. Exposure times were limited to 900 s with a positional accuracy of $<$ 1$^{\prime\prime}$.
Data reduction included bias subtraction and flat-field correction using the IRAF software
package. Extraction (task {\tt extractor}), including background sky subtraction, of data
allowed the creation of one--dimensional spectra, wavelength calibrated using standard lines
from a Cu-Ar arc. Observing conditions were not photometric, seeing was limited to 2$^{\prime\prime}$
at best but varied throughout the evening. We show in Fig.~\ref{Halpha} the resulting spectrum: 
its moderate signal--to--noise ratio does not allow us to make a spectral classification, 
but a significant H$_{\alpha}$ emission line, with equivalent width EW = --37.44 \AA,~is clearly visible. 

Despite no periodicities being found in the OGLE light curve of this star,
these findings indicate that source \# 13 is a Be HMXRB. Since sources of this class
typically have a transient nature, we searched for hard X-ray outbursts in all the public data obtained with the \textit{INTEGRAL} 
satellite \citep{Winkler2003}, that is
all pointings obtained with the IBIS instrument \citep{Ubertini2003} in which the source position was 
within 12\,degrees  from the center of the field of view. This led to a total of 619 pointings,  with different
exposure times from about 1800\,sec to about 3600\,sec, spanning from July 2003 to June 2009.
We used version 9.0 of the Off-line Scientific Analysis software to analyse the data. For each pointing we extracted an image in
the 17--40 keV energy band, but in none of them was the source significantly detected.
The typical flux upper limit of the individual images is on the order of $\sim$20 mCrab. 
We also produced a deep image by summing all the individal pointings, reaching a total exposure of  about 2\,Msec.
The source was not detected, with a  5 $\sigma$ upper limit of about 1 mCrab in the 17--40 keV range, which is well above the 
extrapolation of the flux seen by \EPIC.

\section{Discussion}\label{sec:6} 

We have detected 30 sources with hardness ratios indicating a
column density N$_{\rm H} > 3 \times 10^{23}$ cm$^{-2}$.
Five of these sources are HMXRBs, of which one was discovered in our search and spectroscopically
identified as a new Be binary system.
Most of the remaining 25 sources have very faint (V$>$18) or no optical
counterparts, with colors not compatible with reddened early type-stars 
(unless a significant IR excess is present). 
Therefore, they can hardly be considered as candidate absorbed HMXRBs.

Spectral information for a few of the brightest sources of our sample was previously reported in 
other works. The \XMM\ spectrum of the 25 s pulsar XMMU J004814.1--731003 was closely fit with a power law of photon 
index 1.33$\pm$0.27 and N$_{H} = 5\times10^{22}$  cm$^{-2}$, confirming  our conclusion that it is a highly absorbed source. 
Our brightest source (\# 18) is  the 293 s  pulsar RX J0058.2--7231, for which \citet{Haberl2008a}  reported an 
absorption N$_{H} = (1.35\pm0.32)\times10^{22}$ cm$^{-2}$, considerably smaller than the value we inferred from the hardness ratio. 
This might be due to the results of \citet{Haberl2008a} being obtained from the June 2007 \XMM\ observation, 
while our results for this source are based on the SMC Survey pointing carried out in October 2009, when the source was about a 
factor of ten brighter. HMXRBs can display large variations in their intrinsic absorption, as shown for example in the case 
of the long period (1323 s) SMC pulsar RXJ0103.6--7201 \citep{Eger2008}.

We do not expect of course all the sources in our sample to be HMXRBs in the SMC.
A large fraction of them will turn out to be background-absorbed AGNs.
On the other, hand our sample should not contain foreground 
stars belonging to our Galaxy, which would show softer and less absorbed X-ray spectra. 

A rough estimate of the number of absorbed AGNs in our sample can be derived using published
number flux relations for extragalactic sources.   
For example, based on the LogN-LogS relations for AGNs with N$_{H} = 3\times10^{23}$ cm$^{-2}$
by \citet{Gilli2007},
we obtain a surface density of $\sim$4  AGNs per square degree at a
flux limit of $4\times 10^{-14}$ erg cm$^{-2}$ s$^{-1}$.
Our survey  covered a sky area of $\sim$5 square degrees, although with a non-uniform sensitivity,
because of telescope vignetting and different exposure times.  
Owing to the uncertainties in the LogN-LogS relations and the small number of objects, 
this prediction is consistent with the number of unidentified sources in our sample.

\section{Conclusions}\label{sec:7}

Motivated by the presence of several highly absorbed HMXRBs in our Galaxy, we have carried out
a search for similar systems in the \SMC\, exploiting the nearly complete coverage of this galaxy
obtained with \XMM.  Our selection criteria, corresponding to an absorption threshold of 
$\sim3\times10^{23}$ cm$^{-2}$, were met by four known HMXRBs and led to the 
discovery of a new highly absorbed Be binary in the SMC. 
We also selected other 25 additional sources, among which other highly absorbed HMXRBs might be present.
Since we also expect a good fraction of these sources to be AGNs, we can say that the 
SMC does not contain much more than $\sim$10 persistent HMXRBs with 
intrinsic N$_H>3\times10^{23}$ cm$^{-2}$.   
We cannot exclude the presence of a much larger population of intrinsically absorbed binaries
of transient nature, which could not be detected in our survey if their quiescent luminosity is 
below a few 10$^{34}$ erg  s$^{-1}$.

\section*{Acknowledgments}
\begin{acknowledgements}
We wish to thank Lara Sidoli for the useful discussion.
This work is based on observations obtained with \XMM, an ESA science
mission with instruments and contributions directly funded by ESA
Member States and NASA. The \XMM~data analysis is supported by the
Italian Space Agency (ASI) (ASI/INAF contract I/032/10/0).
We were granted observation time at the South African Astronomical
Observatory (SAAO) and wish to thank them for their kind help and
accommodations.
Travel to the SAAO was funded by Australian Government AINSTO AMNRF grant
number 10/11-O-06.
Partly based on observations with \textit{INTEGRAL}, an ESA project
with instruments and science data centre funded by ESA member states
(especially the PI countries: Denmark, France, Germany, Italy,
Spain, and Switzerland), Czech Republic and Poland, and with the
participation of Russia and the USA.
IRAF is the Image Reduction and Analysis Facility, a general purpose software system
for the reduction and analysis of astronomical data. IRAF is written and supported by the
IRAF programming group at the National Optical Astronomy Observatories (NOAO) in
Tucson, Arizona. NOAO is operated by the Association of Universities for Research in
Astronomy (AURA), Inc. under cooperative agreement with the National Science
Foundation.
Extractor is part of an IRAF package (PNDR) developed to assist with the efficient
reduction of long-slit spectra of emission line objects. It was developed as part of the
Macquarie/AAO/Strasbourg Halpha (MASH) Planetary Nebulae Catalogue.
\end{acknowledgements}

\bibliographystyle{aa}
\bibliography{biblio}

\end{document}